\documentclass[prl,twocolumn,showpacs,preprintnumbers,amsmath,amssymb]{revtex4}

\usepackage{graphicx}
\usepackage{dcolumn}
\usepackage{bm}

\begin{document}

%\preprint{APS/123-QED}

\title{Normal Heat Conduction in a Chain with Weak Interparticle Anharmonic Potential}

\author{Emmanuel Pereira}
 \email{emmanuel@fisica.ufmg.br}
\author{Ricardo Falcao}
 \email{rfalcao@fisica.ufmg.br}
\affiliation{Departamento de F\'{\i}sica-ICEx, UFMG, CP 702,
30.161-970 Belo Horizonte MG, Brazil }

\date{\today}

\begin{abstract}
We analytically study heat conduction in a chain with interparticle
interaction $V(x)=\lambda[1-\cos(x)]$ and  harmonic on-site
potential. We start with each site of the system connected to a
Langevin heat bath, and investigate the case of small coupling for
the interior sites in order to understand the behavior of the system
with thermal reservoirs at the boundaries only. We study, in a
perturbative analysis, the heat current in the steady state of the
one-dimensional system with weak interparticle potential. We obtain
an expression for the thermal conductivity, compare the low and high
temperature regimes, and show that, as we turn off the couplings
with the interior heat baths, there is a ``phase transition:'' the
Fourier's law holds only at high temperatures.
\end{abstract}

\maketitle

The understanding of heat conduction in a lattice system of
interacting particles has become a challenging problem of
statistical physics, even in the $1D$ context  \cite{LLP}. A central
issue is finding a model Hamiltonian system for which Fourier's law
holds.  One of the first works in this subject was the (rigorous)
study of the harmonic chain of interacting oscillators coupled to
heat baths at the boundaries \cite{RLL}. The authors show  that the
heat current is independent of the length of the chain, i.e., the
Fourier's law does not hold. Since then, many (very often
conflicting) works have been devoted to the problem, in particular,
to investigations on the effects of nonlinearity and external
potentials in the behavior of the heat current. We recall some
results. In \cite{PC}, the authors show that the conductivity is
anomalous (i.e., it diverges) in any one-dimensional momentum
conserving system, but in \cite{GS-PRL} and \cite{GLP-PRL} a
momentum conserving system with finite conductivity is presented. In
\cite{TB}, the authors claim  that the anharmonicity of the on-site
potential is a sufficient condition for a finite thermal
conductivity, but in \cite{SG}, it is shown to be wrong.  Almost all
the results are obtained by means of computer simulations, and, as
emphasized in \cite{D-PRL1}, besides the difficulty to arrive at
correct conclusions from numerical studies, several works use the
Green-Kubo formula for the conductivity, a formula which has never
been rigorously established for this context. In short, more
accurate studies are necessary.

In this scenario, the harmonic Hamiltonian chain of oscillators has
been revisited quite recently \cite{BLL}, but for the case of each
site connected to a thermal reservoir. The steady state is
rigorously computed in the ``self-consistent'' condition, which
means with no heat flow between an inner site and its reservoir. In
such a model, the Fourier's law holds.

The present paper is addressed to the following issues: (i) the
development of  new analytical methods of modeling the heat
conduction problem; (ii) the search for a system with normal
conductivity and with, say, a small anharmonic potential (the
problem which inspired the first investigation of Fermi, Pasta and
Ulam \cite{FPU}); (iii) the understanding of the temperature role on
the thermal conductivity of chains with soft anharmonicity such as
$V = 1-\cos(q_{i+1}-q_{i})$ (there is a recent debate, with
positions against \cite{Hu2005} and in favor of \cite{GS2005} a
phase transition in the rotor model - i.e. finite thermal
conductivity for large $T$, and infinite one for small $T$). Here,
we extend the approach and techniques previously developed in
\cite{PF} in order to treat a chain with thermal reservoirs at the
boundaries only: now we consider different coupling constants among
reservoirs and sites, and investigate the limit of the coupling with
the interior heat bath taken to zero. Our approach is quite general,
but we focus  on the case of a chain of oscillators with a harmonic
on-site potential and interparticle interaction $V = \lambda
[1-\cos(q_{i+1}-q_{i})]$. We obtain (in a perturbative analysis) an
expression for the thermal conductivity and investigate the
Fourier's law: for our model, as we turn off the the couplings
between inner sites and their reservoirs, it holds only at high
temperatures.

Now we introduce the model. We consider the  Langevin dynamics of an
anharmonic crystal with stochastic heat bath at each site.
Precisely, we start from $N$ oscillators with Hamiltonian
\begin{equation}
H(q,p)=\sum_{j=1}^{N}\frac{1}{2}\left [p_j^2+Mq_j^2\right ] +
\frac{1}{2}\sum_{j\neq l=1}^{N}\lambda[1 - \cos(q_l - q_j)] ,
\label{Hamiltonian}
\end{equation}
($d=1$ and next-neighbor interactions are assumed later) where
$M>0$, with time evolution, for $j=1,\ldots,N $ ,
\begin{equation}
dq_j = p_jdt; ~~ dp_j = -\frac{\partial H}{\partial q_j}dt-\zeta_{j}
p_jdt+\gamma^{1/2}_jdB_j ;  \label{eqdynamics}
\end{equation}
where $B_j$ are independent Wiener processes; $\zeta_{j}$ is the
heat bath coupling for the $j^{th}$ site; and $\gamma_j=2\zeta_{j}
T_j$, where $T_j$ is the temperature of the  $j^{th}$ heat bath.

As usual, we define the energy of the oscillator $j$ as
\begin{equation}
H_j(q,p)=\frac{1}{2}p_j^2+U^{(1)}(q_j)+\frac{1}{2}\sum_{l\neq
j}U^{(2)}(q_j-q_l) , \label{lenergy}
\end{equation}
where the expression for $U^{(1)}$ and $U^{(2)}$ follow immediately
from (\ref{Hamiltonian}) and $\sum_{j=1}^{N}H_j=H$. Then, we get
\begin{equation}
\left < \frac{dH_j(t)}{dt} \right > =\left < R_j(t) \right > - \left
<\mathcal{F}_{j\rightarrow}-\mathcal{F}_{\rightarrow j} \right >,
\label{venergy}
\end{equation}
where $\left <\cdot\right>$ denotes the expectation with respect to
the noise distribution, and
\begin{equation}
\left <R_j(t)\right >=\zeta_{j}\left ( T_j- \left <p_j^2\right
> \right)\label{reser}
\end{equation}
gives the energy flux from the $j^{th}$ reservoir to the $j^{th}$
site. The remaining terms are related to the energy current inside
the system and they are  given by
\begin{equation}
\mathcal{F}_{j\rightarrow} = \sum_{l>j}\nabla U^{(2)}\left (
q_j-q_l\right )\frac{p_j+p_l}{2}; \label{flux}
\end{equation}
$\mathcal{F}_{j\rightarrow}$ describes the heat flow from the
$j^{th}$ to the $l^{th}$ sites; $\mathcal{F}_{\rightarrow j}$ is
obtained from the formula for  $\mathcal{F}_{j\rightarrow}$ by
changing $l$ with $j$. It is useful to introduce the phase-space
vector $\phi=(q,p)$ with $2N$ coordinates and write the equation for
the dynamics (2) as
\begin{equation}
\dot{\phi}= -A\phi- {U^{(2)}}' + \sigma\eta, \label{dynamics}
\end{equation}
where $A$ and $\sigma$ are $2N\times 2N$ matrices given by
\begin{eqnarray}
A=\left (
\begin{array}{cc}
0 & -I \\
\mathcal{M} & \Gamma \end{array} \right ),  & \sigma=\left (
\begin{array}{cc}
0 & 0 \\
0 & \sqrt{2\Gamma\mathcal{T}}
\end{array}\right ).\label{defin}
\end{eqnarray}
$I$ above is the unit $N\times N$ matrix;  and
$\mathcal{M}$,$\Gamma$,$\mathcal{T}$ are diagonal $N \times N$
matrices: $\mathcal{M}_{jl}=M\delta_{jl}$ ,
$\Gamma_{jl}=\zeta_{j}\delta_{jl}$ ,
$\mathcal{T}_{jl}=T_j\delta_{jl}$. $\eta$ are independent
white-noises; ${U^{(2)}}'$  is the derivative of the $U^{(2)}$ term
in $H$ in relation to $q$ (note that its contribution to
$\dot{\phi}_{k}$ is nonzero only for $k > N$).

To study the dynamics we adopt the following strategy. First, we
consider the system  with $U^{(2)}=0$, and stay with $N$ independent
sites connected, each one, to a heat bath. To recover the original
dynamical system, we introduce the interaction among the sites and
calculate the changes using techniques of stochastic differential
equations. The solution of (\ref{dynamics}) above with
$U^{(2)}\equiv0$, is the Ornstein-Uhlenbeck process
\begin{equation*}
\phi(t)=e^{-tA}\phi(0)+\int_{0}^{t}ds ~~ e^{-(t-s)A}\sigma\eta(s) .
\label{Orns}
\end{equation*}
For simplicity we take $\phi(0)=0$. The covariance of this Gaussian
process is
\begin{eqnarray}
\left < \phi(t)\phi(s)\right >_{0} &\equiv & \mathcal{C}(t,s)=\left
\{
\begin{array}{c}
e^{-(t-s)A}\mathcal{C}(s,s) \quad t\geq s, \\
\mathcal{C}(t,t)e^{-(s-t)A^{T}} \quad t\leq s,
\end{array} \right . \\
&& \mathcal{C}(t,t)=\int_{0}^{t}ds ~~ e^{-sA}\sigma^{2} e^{-sA^{T}}.
\nonumber \label{covariance}
\end{eqnarray}
From an easy computation (e.g. diagonalizing $A$), it follows that
(for a single site $\phi_{j}$)
\begin{eqnarray}
\exp\left (-tA \right
)=e^{-t\frac{\zeta_{j}}{2}}\cosh(t\rho_{j})\left \{ \left (
\begin{array}{cc}
1 & 0 \\
0 & 1
\end{array} \right ) \right .\nonumber  \\
\left . +\frac{\tanh(t\rho_{j})}{\rho_{j}}\left (
\begin{array}{cc}
\frac{\zeta_{j}}{2} & 1 \\
-M & -\frac{\zeta_{j}}{2}
\end{array} \right ) \right \},
\end{eqnarray}
$\rho_{j}=\left ( (\zeta_{j}/2)^2-M\right )^{1/2}$; the expressions
for $\phi$ involving $2N\times 2N$ matrices are immediate. We assume
that $\zeta_{j}/2,~~ M>0$. If $\left(\zeta_{j}/2\right)^{2}>M$, then
$\rho_{j}$ is real; otherwise, $\rho_{j}$ is pure imaginary, but it
does not spoil the dynamics: $\cosh(t\rho)$ in the formula above
becomes $\cosh(t\cdot i\rho') = \cos(t\rho')$, etc. In this case
(i.e., with $U^{(2)}=0$), as $t\rightarrow\infty$ we have a
convergence to equilibrium  and the stationary state is Gaussian,
with mean zero and covariance
\begin{eqnarray}
C=\int_{0}^{\infty}ds ~~ e^{-sA}\sigma^{2} e^{-sA^{T}}=\left (
\begin{array}{cc}
\frac{\mathcal{T}}{M} & 0 \\
0 & \mathcal{T}
\end{array} \right ),\label{covariances}
\end{eqnarray}
where $\mathcal{T}$ is a diagonal matrix with elements
$T_i\delta_{ij}$.
To introduce the anharmonic coupling potential, we
use the Girsanov theorem \cite{O}, which establishes a measure $\nu$
for the complete process (\ref{dynamics}) as an integral
representation involving the measure $\mu_{\mathcal{C}}$ associated
to the process without the potential $U^{(2)}$. Precisely, for any
measurable set $R$, it states that $\rho(R)=E_0(1_{R}Z(t))$, where
$E_{0}$ is the expectation for $\mu_{\mathcal{C}}$ (the measure for
the process with $U^{(2)}= 0$); $1_{R}$ denotes the characteristic
function, and
\begin{eqnarray}
&Z(t) = \exp\left(\int_{0}^{t}u\cdot dB
-\frac{1}{2}\int_{0}^{t}u^2ds\right);& \nonumber \\
&\gamma_i^{1/2}u_i = -\nabla_{i-N}U^{(2)} ;& \label{girs}
\end{eqnarray}
the inner products above are in $\mathbb{R}^{2N}$ and $\nabla_{k}$
means the derivative in relation to $\phi_{k}$. From (\ref{defin})
and the expression above for $u_i$, we have that  $u_i$ is
nonvanishing only for $i>N$ (i.e., $i\in [N+1,N+2,\ldots,2N ]$). In
what follows we will use the index notation: $i$ for index values in
the set $[N+1,N+2,\ldots,2N]$, $j$ for  values in the set
$[1,2,\ldots,N]$, and $k$ for values in $[1,2,\ldots,2N]$. We will
also be restricted to next-neighbor interactions, i.e., we take
\begin{eqnarray*}
U^{(2)} = \frac{1}{2}U^{(2)}(\phi_{1}-\phi_{2}) + \frac{1}{2}U^{(2)}(\phi_{N-1}-\phi_{N}) + \\
+\sum_{j=2}^{N-1}\frac{1}{2}\left\{U^{(2)}(\phi_{j-1}-\phi_{j})+
U^{(2)}(\phi_{j}-\phi_{j+1})\right\},
\end{eqnarray*}
where $U^{(2)}(x) = \lambda[1-\cos(x)]$. Using the $i,j,k$ index
notation above, we may rewrite the stochastic equations for the
decoupled process (where $U^{(2)}=0$) as
\begin{equation}
d\phi_j = -A_{jk}\phi_kdt; \quad
d\phi_i=-A_{ik}\phi_kdt+\gamma_i^{1/2}dB_i; \label{stoeq}
\end{equation}
the sum over $k$ (in $[1,2\ldots,2N])$ is assumed above (as well as
obvious sum over some indices in what follows). Now, let us make
explicit the terms in $Z(t)$. We have
\begin{equation*}
u_idB_i=\gamma_i^{-1/2}u_i\gamma_i^{-1/2}dB_i =
-\gamma_i^{-1}U'_{i-N} \left (d\phi_i+A_{ik}\phi_kdt\right ) ,
\end{equation*}
where we dropped out the upper index $^{(2)}$ in $U^{(2)}$ above
(and in what follows); $U'_{i-N}$ means the derivative in relation
to $\phi_{i-N}$, i.e., $ U'_{j} = U'(\phi_{j}-\phi_{j+1}) -
U'(\phi_{j-1}-\phi_{j}); ~~~~ U'(\phi_{j}) = -\lambda\sin(\phi_{j}).
$ From It\^o formula \cite{O}, it follows that
\begin{eqnarray*}
&-\left(U'_{j,j+1}-U'_{j-1,j}\right)\frac{d\phi_i}{\gamma_i}= -dF+&
\nonumber \\
&+\frac{\phi_i dt}{\gamma_i}\left(U''_{j,j+1}[\phi_{j+N}-
\phi_{j+N+1}] - \nonumber \right.\left. U''_{j,j-1}[\phi_{j+N-1}
-\phi_{j+N}]\right),&\\
&F(\phi) = \gamma_i^{-1}\left(U'_{j,j+1} -
U'_{j-1,j}\right)\phi_{i}, &
\end{eqnarray*}
where $U'_{j,j+1} \equiv U'(\phi_{j} - \phi_{j+1})$, etc; and
$j=i-N$ in the expressions above. Hence, we get
\begin{eqnarray}
&\int_{0}^{t} u_{i}dB_{i} = -\gamma_{i}^{-1}\Big (
[U'_{j,j+1} - U'_{j-1,j}][\phi_{i}(t) -\phi_{i}(0)] +&\nonumber \\
&+\int_{0}^{t} \phi_{i}(s)[U''_{j,j+1}(\phi_{i}-\phi_{i+1})-U''_{j-1,j}(\phi_{i-1}-\phi_{i})](s) ds + &\nonumber \\
& \int_{0}^{t}[U'_{j,j+1} - U'_{j-1,j}]A_{ik}\phi_{k}(s) ds \Big),&
\label{udB},
\end{eqnarray}
and, for the $u^{2}$ term,
\begin{equation}
- \frac{1}{2}\int_{0}^{t} u_{i}^{2} ds =
-\frac{\gamma_{i}^{-1}}{2}\int_{0}^{t}[U'_{j,j+1} -
 U'_{j-1,j}]^{2}(s) ds ,
\end{equation}
where, again, $j=i-N$ an the sum over $i$ (and so $j$) is assumed
above. Then, for the correlation functions, we obtain an integral
representation involving a ``perturbative'' potential and a Gaussian
measure. E.g., for the two-point function we get $\left <
\phi_k(t_1)\phi_q(t_2) \right
>= {\cal N}\int\phi_k(t_1)\phi_q(t_2)Z(t)d\mu_{\mathcal{C}}(\phi)$, $t_{1},t_{2}<t$;
where $Z(t)=e^{-W}$, $W$ is described by the several terms presented
by the expressions above; ${\cal N}$ is the normalization.

The heat flow in the steady state is related to the formula
(\ref{flux}). Precisely, for the case of next-neighbor interactions,
the average over the stationary distribution for the current ${\cal
F}_{v\rightarrow v+1}$ is obtained as the limit
\begin{eqnarray*}
&\lim_{t\rightarrow\infty} \langle {\cal F}_{v\rightarrow
v+1}\rangle =\lim_{t\rightarrow\infty} & \\ &\left.\int
[U'_{v,v+1}(\phi_{u} +
\phi_{u+1})/2](t)Z(t)d\mu_{\mathcal{C}}(\phi)\right/\int
Z(t)d\mu_{\mathcal{C}}(\phi)& ,
\end{eqnarray*}
where $v=u-N$ ($u>N$, obviously). Now, we carry out the computation.
First note that $\mathcal{C}(t,s)$, given by
(\ref{covariance})-(\ref{covariances}), may be written as (for
$t>s$) $ \mathcal{C}(t,s)=\exp(-(t-s)A)C+\mathcal{O}\left(
\exp[-(t+s)\zeta/2]\right )$, and the effects of the second term in
the r.h.s of the equation above disappear in the correlation formula
in the limit of $t\rightarrow\infty$ (recall that we must take this
limit $t\rightarrow\infty$ in order to reach the steady state).
 Writing $Z(t) = e^{-\lambda
W}$, the previous formula becomes
\begin{equation*}
\lim_{t\rightarrow\infty} \langle {\cal F}_{v\rightarrow v+1}\rangle
= \lim_{t\rightarrow\infty} \int \Omega e^{-\lambda W}
d\mu_{\mathcal{C}}(\phi)\left/\int e^{-\lambda W}
d\mu_{\mathcal{C}}(\phi)\right. ,
\end{equation*}
$\Omega$ given by the product of $U'$ and $\phi$ described above. Up
to first order in $\lambda$ (i.e., for weak interaction between two
sites), we have $ \langle {\cal F}_{v\rightarrow v+1}\rangle =
\langle \Omega \rangle_{\cal C} + \langle \Omega ; -\lambda W
\rangle_{\cal C}$, where $\langle \cdot \rangle_{\cal C}$ means the
average in respect to $d\mu_{\cal C}$; $\langle \cdot ; \cdot
\rangle_{\cal C}$ means the truncated expectation. It is easy to see
that $\langle \Omega \rangle_{\cal C}$ vanishes: $U'$ depends on
$\phi_{v}$ or $\phi_{v+1}$, and $\langle
\phi_{v}(t)\phi_{u}(t)\rangle_{\cal C} = {\cal C}_{v,u}(t,t) = 0$
(for any $v\leq N$ and $u>N$). The terms in $-\lambda W$ are given
by those describing $\int_{0}^{t}u_{i}dB_{i}$ (\ref{udB}),
discarding that one involving $\phi(0)$  which vanishes in the
computation as $t\rightarrow\infty$ (${\cal C}(t,0)\rightarrow 0$ as
$t\rightarrow\infty$). Note that we stay involved with expressions
such as
\begin{eqnarray*}
&\langle U'_{v,v+1}\phi_{u}(t); U'_{j,j+1}\phi_{\tilde{j}}(s)
\rangle_{\cal C} = &\\
& -\frac{1}{4}\left< e^{i\phi_{v+1}(t)}e^{-i\phi_{v}(t)} \phi_{u}(t)
; e^{i\phi_{j+1}(s)}e^{-i\phi_{j}(s)}
\phi_{\tilde{j}}(s)\right>_{\cal C} + \ldots&
\end{eqnarray*}
i.e., with integrals like $ \int e^{i(h_{1} + \ldots +
h_{6})\cdot\phi} d\mu_{\cal C} - \int e^{i(h_{1} + h_{2} +
h_{3})\cdot\phi} d\mu_{\cal C}\int e^{i(h_{4} + h_{5} +
h_{6})\cdot\phi} d\mu_{\cal C}$, ($\phi_{u}(t)$ is obtained from
$e^{ih\cdot\phi}$, obviously, by taking the derivative in relation
to $h_{u}(t)$ and making $h\equiv 0$; the same for
$\phi_{\tilde{j}}(s)$). After these integrations in $\phi$, we get
expressions such as
$$
\lim_{t\rightarrow\infty} \int_{0}^{t} ds ~~ {\cal C}_{u,j+1}(t,s)
{\cal C}_{v+1,\tilde{j}}(t,s) \exp[-\sum_{\alpha,\beta} {\cal
C}_{\alpha,\beta}(t_{\alpha},t_{\beta})/2] ,
$$
where $t_{\alpha}, t_{\beta}$ are $t$ or $s$. For the case of small
temperatures $T_{j}$, we have $\exp[{\cal C}_{\alpha,\beta}]\approx
1$. Thus, after all $\phi$ and $s$ integrations (taking also the
limit $t\rightarrow \infty$), we obtain
\begin{equation}
{\cal F}_{v\rightarrow v+1} =
\frac{-\lambda^{2}}{2M(\zeta_{v}+\zeta_{v+1})}\left[\frac{\zeta_{v}}{\zeta_{v+1}}T_{v+1}
- \frac{\zeta_{v+1}}{\zeta_{v}}T_{v}\right] ,
\end{equation}
(to simplify the notation we write ${\cal F}_{v\rightarrow v+1}$
instead of $\lim_{t\rightarrow \infty}\langle {\cal F}_{v\rightarrow
v+1}\rangle$). For the high temperature regime, we have to deal with
expressions like $ \lim_{t\rightarrow\infty}\int_{0}^{t} f(t,s)
e^{-\theta g(t,s)} ds, $ where $\theta \propto T$. We use the
Laplace method \cite{M} to get their asymptotic behavior
$\theta\rightarrow\infty$. We obtain
\begin{widetext}
\begin{eqnarray}
{\cal F}_{v\rightarrow v+1} & \approx & \frac{\lambda^{2}}{8} \left(
\left(\frac{2}{\zeta_{v+1}}  +
 \frac{1}{\zeta_{v+2}}\right)e^{-(T_{v+2}+T_{v})/2M}
 - \left(\frac{2}{\zeta_{v}}
 + \frac{1}{\zeta_{v-1}}\right)e^{-(T_{v+1}+T_{v-1})/2M}\right)
 + \nonumber \\
 && + \frac{\lambda^{2}}{4(T_{v+1}+T_{v})}\left[\left(\frac{T_{v+1}}{\zeta_{v+1}}
 - \frac{T_{v}}{\zeta_{v}}\right) - \left(\frac{T_{v+1}}{\zeta_{v}}
 - \frac{T_{v}}{\zeta_{v+1}}\right)\right] ;
\end{eqnarray}
\end{widetext}
 with slight changes for the terms with $v=1$ and $v=N$.

The steady state is characterized by $\langle dH_{j}/dt \rangle =0$.
Using this expression and also that
$\lim_{t\rightarrow\infty}\langle \phi^{2}_{i}\rangle = T_{i-N}$
(taking the dominant contribution), which gives
$\lim_{t\rightarrow\infty}\langle R_{j}(t)\rangle = 0$ for the
interior sites $j$ (i.e. the ``self-consistent condition''), we have
$ \mathcal{F}_{1\rightarrow 2}=\mathcal{F}_{2\rightarrow
3}=\ldots=\mathcal{F}_{N-1\rightarrow N} \equiv \mathcal{F}. $
Hence, for $\zeta_{j+1}-\zeta_{j}$ small, summing up ${\cal
F}_{1\rightarrow 2} + {\cal F}_{2\rightarrow 3} + \ldots$ we obtain,
for small temperatures,
$$
\mathcal{F}\cdot
2M(\zeta_{1}+2\zeta_{2}+\cdots+2\zeta_{N-1}+\zeta_{N}) \approx
\lambda^{2}(T_{1}-T_{N}).
$$
For uniform $\zeta$, we have the Fourier's law
$$
\mathcal{F} = \chi(T_{1}-T_{N})/(N-1), ~~~~ \chi=\lambda^{2}/4\zeta
M.
$$
As we make the inner couplings smaller and smaller we lose the
factor $N$ (which comes from $\zeta_{2}+\ldots+\zeta_{N-1}$) and the
Fourier's law does not hold anymore. For the case of high
temperatures the sum of all ${\cal F}_{v\rightarrow v+1}$ gives us,
for $\zeta_{v+1}-\zeta_{v}$ small,
$$
\mathcal{F}(N-1) \approx
\frac{\lambda^{2}}{4}\left(\frac{e^{-T_{N}/M}}{\zeta_{N}} -
\frac{e^{-T_{1}/M}}{\zeta_{1}}\right),
$$
which (essentially) does not depend on inner heat bath couplings,
and so, the Fourier's law still holds when we make them smaller and
smaller. Taking $\zeta_{N}=\zeta_{1}=\zeta$, and $T_{N} = T_{1} +
\delta$ ($\delta$ small) the expression above becomes
\begin{equation}
\mathcal{F} \approx
\frac{\lambda^{2}}{4M\zeta}e^{-T/M}\frac{(T_{1}-T_{N})}{(N-1)},
\end{equation}
where $e^{-T/M}=-[e^{-(T_{1} + \delta)/M}- e^{-T_{1}/M}]/[(T_{1} +
\delta)/M - T_{1}/M]$, i.e., the conductivity decays exponentially
at high temperatures, as in the rotor model \cite{GS-PRL}. Thus,
still concerning the chain of rotators and the recent debate about
the existence of a phase transition  \cite{Hu2005}, \cite{GS2005},
based on our results with cosine interactions, we believe in a
divergent conductivity in low temperatures, as claimed in
\cite{GS2005}.

To argue about the reliability of our treatment, we recall some
previous related works where the perturbative analysis gives the
same result of the rigorous treatment. For the simpler case of the
harmonic chain of oscillators with a bath at each site and identical
next-neighbor interactions, a first order perturbative analysis
\cite{PF} (for weak interactions, in a similar approach to that
described here) gives the same result of the complete and rigorous
treatment \cite{BLL}. And following the procedures described here
(analyzing different $\zeta_{j}$), one may see that the perturbative
result for this harmonic chain, in the limit of zero coupling
between reservoirs and inner sites, will lead to the rigorous result
obtained for the harmonic chain with thermal baths at the boundaries
\cite{RLL}. We still recall some previous works considering
nonconservative stochastic Langevin systems (in contact with thermal
reservoirs at the same temperature), but involving similar integral
expressions for the correlations. There, the time decay of the two
and/or four-point functions is detailed investigated in the regions
of low and high temperatures. For the low temperature regime and
weak interaction among the sites, we rigorously prove \cite{FOPS}
that the complete treatment of the two and four-point functions adds
only small corrections to the perturbative results \cite{SBFP}. For
the same nonconservative system at high temperature, we developed a
 cluster expansion \cite{TPP} which supports the perturbative
analysis \cite{PPRE}.

In short, in the present letter we develop an analytical method of
modeling the heat conduction problem and study the heat current at
the steady state of an anharmonic chain with weak interparticle
(cosine) potential in order to investigate an old problem of heat
conduction: may small anharmonic interactions lead to normal
conductivity? We show that, if we keep the thermal reservoirs at the
boundaries only, the Fourier's law holds for high but not for low
temperatures.

We acknowledge financial support from CNPq (Brazil).

\end{document}